\documentclass[aps,superscriptaddress,nofootinbib,twocolumn]{revtex4-1}

\usepackage{mathrsfs}
\usepackage{amsfonts}
\usepackage{amssymb}
\usepackage{amsmath}
\usepackage{amsthm}
\usepackage{graphicx}
\usepackage[usenames,dvipsnames]{color}
\usepackage[colorlinks=true,citecolor=blue,linkcolor=magenta]{hyperref}
\usepackage{ulem}
\usepackage{lmodern}
\usepackage{microtype}
\usepackage{braket}
\usepackage{booktabs}
\usepackage{pifont}
\usepackage{bbold}
\usepackage{mathtools}
\usepackage{xcolor}
\usepackage{dsfont}
\usepackage{colortbl}
\usepackage{algpascal}
\usepackage{algorithm} 
\usepackage{algpseudocode}
\usepackage{chngcntr}
\usepackage{multirow}
\usepackage{float} 

\newtheorem{theorem}{Theorem}

\begin{document}
\date{\today}

\title{Experimental learning of quantum states}

\author{Andrea Rocchetto}
\email[E-mail:]{andrea.rocchetto@spc.ox.ac.uk}
\affiliation{Department of Computer Science, University College London, London, UK}
\affiliation{Department of Materials, University of Oxford, Oxford, UK}
\author{Scott Aaronson}
\affiliation{Department of Computer Science, University of Texas at Austin, Austin, USA}
\author{Simone Severini}
\affiliation{Department of Computer Science, University College London, London, UK}
\affiliation{Institute of Natural Sciences, Shanghai Jiao Tong University, Shanghai, China}
\author{Gonzalo Carvacho}
\affiliation{Dipartimento di Fisica - Sapienza Universit\`a di Roma, Rome, Italy}
\author{Davide Poderini}
\affiliation{Dipartimento di Fisica - Sapienza Universit\`a di Roma, Rome, Italy}
\author{Iris Agresti}
\affiliation{Dipartimento di Fisica - Sapienza Universit\`a di Roma, Rome, Italy}
\author{Marco Bentivegna}
\affiliation{Dipartimento di Fisica - Sapienza Universit\`a di Roma, Rome, Italy}
\author{Fabio Sciarrino}
\email[E-mail:]{fabio.sciarrino@uniroma1.it}
\affiliation{Dipartimento di Fisica - Sapienza Universit\`a di Roma, Rome, Italy}

\begin{abstract}
The number of parameters describing a quantum state is well known to grow
exponentially with the number of particles. This scaling clearly limits our
ability to do tomography to systems with no more than a few qubits and has
been used to argue against the universal validity of quantum mechanics itself.
However, from a computational learning theory perspective, it can be shown that,
in a probabilistic setting, quantum states can be approximately learned using
only a linear number of measurements. Here we experimentally demonstrate this
linear scaling in optical systems with up to $6$ qubits. Our results highlight
the power of computational learning theory to investigate quantum information,
provide the first experimental demonstration that quantum states can be
``probably approximately learned'' with access to a number of copies of the
state that scales linearly with the number of qubits, and pave the way to
probing quantum states at new, larger scales.
\end{abstract}

\maketitle

The exponential scaling of the wavefunction, arising from the tensor product description of multi-particle states, is one of the remarkable properties of quantum systems. If exploited correctly it can be used to achieve the computational advantages theorised in quantum information processing, but it can also lead one to question the consistency of quantum mechanics itself: does it make sense at all to talk about objects with more parameters than the number of atoms in the universe?

One of the problems arising from the exponential scaling of the wavefunction can be formalised in quantum tomography~\cite{banaszek2013focus, haah2017sample, gross2010quantum, quantumTomo, TomoHighDim, lvovsky2009continuous, cramer2010efficient, o2016efficient}. The central task of quantum tomography is to produce a description of an $n$-qubit state given the ability to prepare and measure $k$ of its copies~\cite{lvovsky2009continuous}. Characterising an unknown quantum state is a fundamental tool in quantum information processing. A survey of the major applications and present challenges in state tomography can be found in the review by Banaszek, Cramer, and Gross~\cite{banaszek2013focus}. State estimation is, in general, an expensive procedure. For an $n$-qubit quantum state it can be shown that estimating the ideal state up to an approximation parameter $\epsilon$ requires $\Omega(4^n / \epsilon^2)$ operations~\cite{haah2017sample}. Although prior information, such as the state being low-rank, can be used to reduce the computational cost of quantum tomography~\cite{gross2010quantum, cramer2010efficient, o2016efficient}, there is no hope of overcoming the exponential scaling for general unknown quantum states. Given this difficulty, it is valuable to interpret quantum tomography as a learning problem, with the hope of using the well-developed machinery of computational learning theory, for optimizing the number of required measurements.

Computational learning theory \cite{kearns1994introduction, shalev2014understanding} is a research field devoted to studying the design and analysis of machine learning algorithms. Particularly relevant for our purposes is supervised machine learning. Here the learner is presented with a number of examples consisting of input-output pairs and is subsequently assigned the task of predicting the output of a new input. This model of learning has been formalised in computational learning theory by Valiant in 1984 \cite{valiant1984theory} with the introduction of the Probably Approximately Correct (PAC) model. This framework provides two indicators of the efficiency of a learner: the sample complexity and the time complexity. The first is the worst-case number of examples it uses to reach some target competency, while the second one is the worst-case running time of the learner. In this article we are concerned with the sample complexity of the problem of learning quantum states.

Quantum state tomography can be rephrased as a learning problem in the following sense. A full tomography requires a complete set of measurements. Consider a learner that by looking at only a few measurements can predict the outcome of any measurement made on the state. It is easy to see that generating this hypothesis is equivalent to reconstructing the density matrix of the state. Because quantum tomography requires an exponentially large number of measurements we might assume that the same applies for the learning problem.

This apparent exponential scaling of the learning problem for quantum states can be interpreted as formalising the objections of quantum mechanics sceptics (for a critical discussion see \cite{aaronson2013quantum}). Indeed, one of the fundamental tasks of science is to come up with hypotheses that, by explaining past observations, let us predict future observations. A theory that requires an exponentially growing number of observations to produce its hypothesis may signal a problem with the theory itself.

Computational learning theory, and in particular the PAC model, can help to
address these conundrums. By analysing quantum tomography from a computational
learning perspective, Aaronson \cite{aaronson2007learnability} proved that
quantum states can be PAC-learned with a linearly scaling training set. Here we
present the first experimental demonstration of such linear scaling. Our
contributions also include developing a testable model for the main theorem
proved in \cite{aaronson2007learnability} and estimating an important scaling constant. We run the experiments on a
photonic platform including up to $6$ qubits.
Our results demonstrate experimentally an important property of quantum states
and highlight the power of computational learning theory in the quantum
information framework.

\begin{figure}
\centering
\includegraphics[scale=0.18]{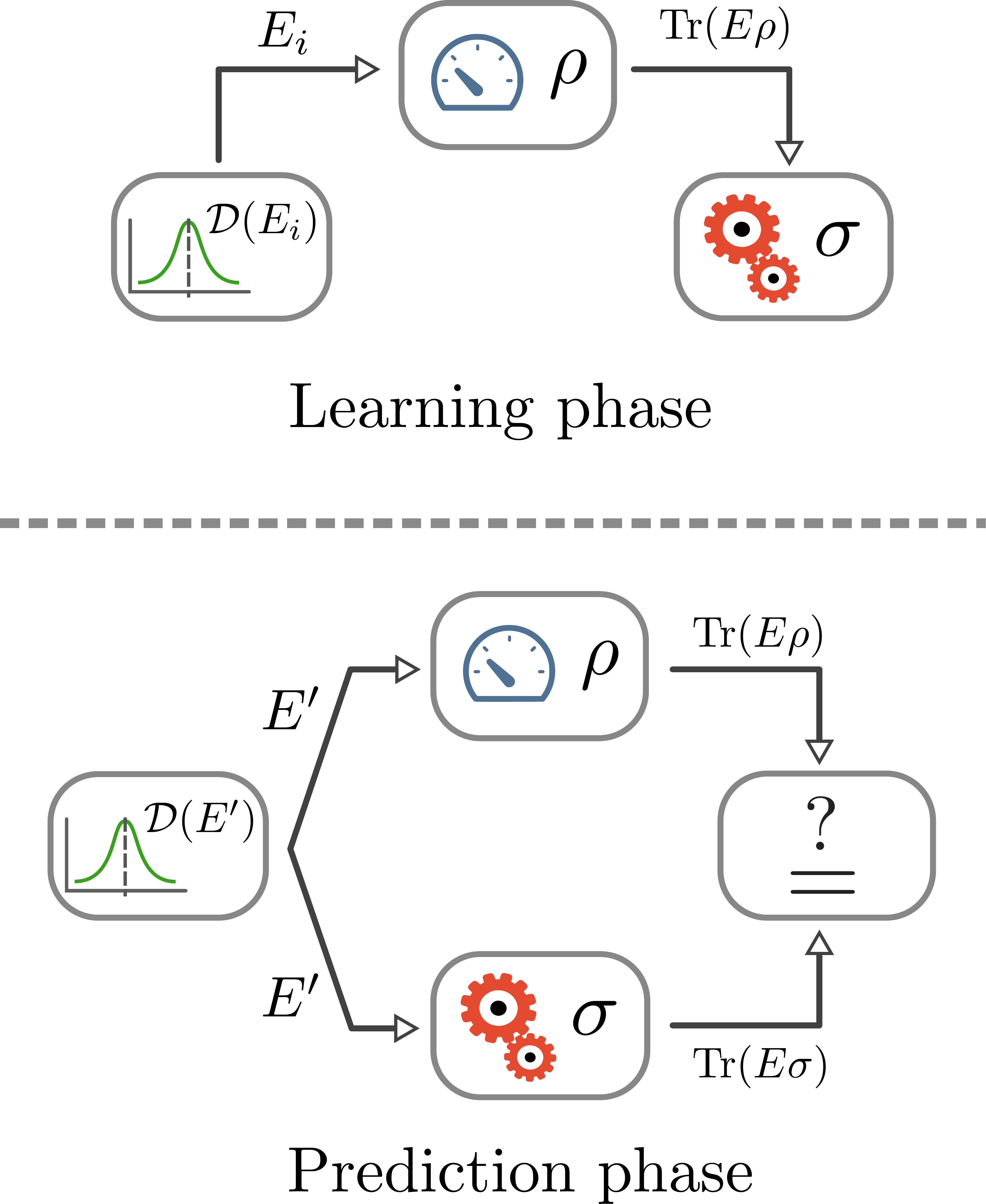}
\caption{\textbf{Schematic of the learning procedure.} In the learning phase (top panel) measurements drawn randomly from $\mathcal{D}$ are performed on the physical state $\rho$. Based on the measurement outcomes the learning algorithm outputs an hypothesis $\sigma$. In the prediction phase (bottom panel) the goal is to predict the experimental outcome of a measurement $E'$ drawn from $\mathcal{D}$ using $\sigma$ as hypothesis. }
\label{fig:learnigSchematic}
\end{figure}

\section*{Quantum learnability theory}\label{sec:learningQstates}

Let us recall some standard definitions in quantum theory. A generic $n$-qubit state $\rho $ is a trace-one, positive semidefinite matrix acting on a Hilbert
space of dimension $2^{n}$. Every observation of
a state is mathematically described by a \textit{positive operator valued measurement} (POVM), $E=\{E^{(j)}\}$, where each $E^{(j)}$ is a Hermitian
positive semidefinite operator such that $\sum_{j}E^{(j)}= I$. The probability of measurement outcome $j$ is $p(j)=$Tr$(E^{(j)}\rho )$. For our purposes, we refer to a measurement of $\rho $ as a \textit{two-outcome POVM} $%
\{E^{(1)}=E,E^{(2)}=1-E\}$ with eigenvalues in $[0,1]$. We denote by $\mathcal{S}$ the set of all measurements on $n$ qubits.

Following Ref.~\cite{aaronson2007learnability} we define the learning of $\rho$ as the task of processing a \textit{training set} composed of $m$ tuples $\{ (E_i, \operatorname*{Tr}(E_i \rho) ) \}$, drawn from a probability distribution~$\mathcal{D}$, in order to predict the ``behaviour'' of $\rho$ on most measurements drawn from~$\mathcal{D}$. This concept of learning is defined in the context of Valiant's PAC model \cite{valiant1984theory}. In this framework, originally developed for Boolean functions but then extended to real-valued ones by Barlett and Long~\cite{barlett1994fat}, a learning algorithm (the learner) tries to approximate with high probability an unknown function $f: \mathcal{X} \rightarrow \mathcal{Y} $ from a training set of random labelled examples. Each labelled example is of the form $(x,f(x))$, where $x$ is distributed according to some unknown distribution~$\mathcal{D}$. In order to make learning possible we restrict the hypothesis that the learner can use to approximate $f$ to a set of functions $\mathcal{H}= \{ h: \mathcal{X} \rightarrow \mathcal{Y} \}$. We refer to $\mathcal{H}$ as the \textit{hypothesis class}. The learning algorithm takes as input the training set and generates a hypothesis $h \in \mathcal{H} $ that approximates $f$. The PAC model makes use of two approximation parameters, $\epsilon$ and $\delta$. The \textit{accuracy parameter} $\epsilon$ determines how far the hypothesis $h$ can be from $f$. The \textit{confidence parameter} gives the probability of sampling a training set that is not representative of the underlying distribution $\mathcal{D}$.  A hypothesis class $\mathcal{H}$ is said to be \textit{PAC-learnable} if there exists an algorithm that, for every probability distribution $\mathcal{D}$ and function $f$ and for every $\epsilon$,$\delta\in (0,1)$, when running the learning algorithm on $m \geq m_{\mathcal{H}}$ examples drawn from $\mathcal{D}$, we have that, with probability at least $1-\delta$, 
$$\Pr_{x\sim \mathcal{D}}[\sigma(x)\neq f(x)]\leq\epsilon.$$
Here by $\sim$ we indicate that $x$ is drawn from~$\mathcal{D}$. The value $m_{\mathcal{H}}$ determines the minimum number of examples required to PAC-learn the class $\mathcal{H}$.  We refer to $m_{\mathcal{H}}$ as the sample complexity of the hypothesis class $\mathcal{H}$. We note that the learner must test the predictions under the same distribution~$\mathcal{D}$ that determines the elements in the training set.

\begin{figure*}[!t]
\includegraphics[width=2\columnwidth]{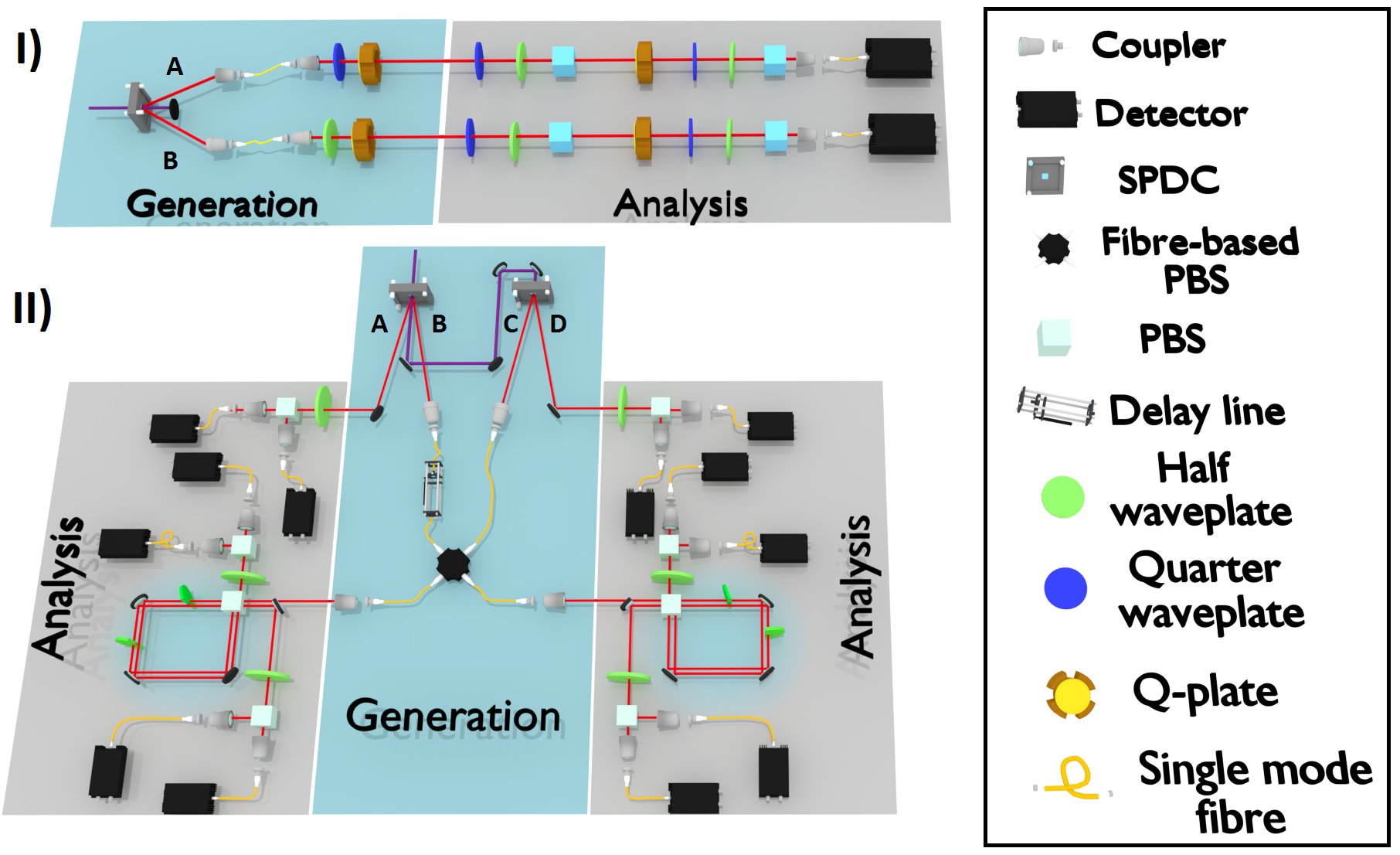}
\caption{ \textbf{Experimental setups for generating the $3$, $4$, $5$ and $6$-qubit GHZ states.} Pictorial representation of the two different experimental setups used to generated the quantum states learned with Theorem~\ref{qoccam}. In setup (I) we makes use of two photons and encode up to $4$ qubits. In setup (II) we makes use of four photons and encode up to $6$ qubits. \textbf{(I)} In the generation stage, the state of each of the two
entangled photons ($1$ and $2$) is locally manipulated via QWPs, HWPs and
q-plates, set to generate a specific GHZ state. The analysis is performed using
QWPs, HWPs and PBSs. The OAM analysis requires a q-plate to transfer the
information encoded in the OAM space to the polarisation degree of freedom which
can be then analysed with standard techniques. After the analysis, both photons
are sent to single mode fibres connected to single photon detectors.
\textbf{(II)} Two polarization-entangled photon pairs are generated via SPDC in
two separated non-linear crystals. Photon $A$ and $D$ of the first and second
pair respectively are sent directly to a HWP and a PBS for polarisation analysis. 
Photons $B$ and $C$ instead are sent to a $50/50$ in-fiber
PBS followed by another PBS which realizes the polarisation-path entanglement.
The two paths go through two HWP and are rejoined in the same PBS forming a
Sagnac-like configuration which allow us to perform polarisation and path analysis
without worrying about phase instabilities.
A motorized delay line is adopted to change the photons wave-packet temporal
overlap in the PBS. The path analysis section is composed by HWP and PBS after which
the photons are coupled into SM fibres connected to single photon detectors.
Generation and analysis sections are represented by cyan and grey zones, respectively.}
\label{fig:lhv}
\end{figure*}

\begin{table*}
\begin{tabular}{*{10}{c}}
\toprule
N of qubits & Experimental Setup & \multicolumn{8}{c}{Photon} \\
& & \multicolumn{2}{c}{A} & \multicolumn{2}{c}{B} & \multicolumn{2}{c}{C} & \multicolumn{2}{c}{D} \\
\midrule
& & $\ket{0}_{1}=\ket{H}$ &   & $\ket{0}_{2}=\ket{H}$ & & & && \\
\multirow{-2}{*}{2} & \multirow{-2}{*}{I}
& $\ket{1}_{1}=\ket{V}$ &   & $\ket{1}_{2}=\ket{V}$ & & & &&\\

\rowcolor[gray]{.9}
& & $\ket{0}_{1}=\ket{R}$ & $\ket{0}_{2}=\ket{+1}$ & $\ket{0}_{3}=\ket{R}$ & & & & \\
\rowcolor[gray]{.9}
\multirow{-2}{*}{3} & \multirow{-2}{*}{I}
& $\ket{1}_{1}=\ket{L}$ & $\ket{1}_{2}=\ket{-1}$ & $\ket{1}_{3}=\ket{L}$ & & & & \\

& & $\ket{0}_{1}=\ket{R}$ & $\ket{0}_{2}=\ket{+1}$ & $\ket{0}_{3}=\ket{R}$
    &$\ket{0}_{4}=\ket{+1}$& & & & \\
\multirow{-2}{*}{4} & \multirow{-2}{*}{I}
& $\ket{1}_{1}=\ket{L}$ & $\ket{1}_{2}=\ket{-1}$ & $\ket{1}_{3}=\ket{L}$
    &$\ket{1}_{4}=\ket{-1}$& & & & \\

\rowcolor[gray]{.9} 
& & $\ket{0}_{1}=\ket{H}$ & & $\ket{0}_{2}=\ket{H}$ & & $\ket{0}_{3}=\ket{H}$ &
    & & \\
\rowcolor[gray]{.9}   
\multirow{-2}{*}{3} & \multirow{-2}{*}{II}
& $\ket{1}_{1}=\ket{V}$ & & $\ket{1}_{2}=\ket{V}$ & & $\ket{1}_{3}=\ket{V}$ &
    & & \\
    
& & $\ket{0}_{1}=\ket{H}$ & & $\ket{0}_{2}=\ket{H}$ & & $\ket{0}_{3}=\ket{H}$ &
    &$\ket{0}_{4}=\ket{H}$ & \\
\multirow{-2}{*}{4} & \multirow{-2}{*}{II}
& $\ket{1}_{1}=\ket{V}$ & & $\ket{1}_{2}=\ket{V}$ & & $\ket{1}_{3}=\ket{V}$ &
    &$\ket{1}_{4}=\ket{V}$ & \\

\rowcolor[gray]{.9}
& & $\ket{0}_{1}=\ket{H}$ & $\ket{0}_{2}=\ket{a}$ & $\ket{0}_{3}=\ket{H}$
    &&$\ket{0}_{4}=\ket{H}$&&$\ket{0}_{5}=\ket{H}$ & \\
\rowcolor[gray]{.9}
\multirow{-2}{*}{5} & \multirow{-2}{*}{II}
& $\ket{1}_{1}=\ket{V}$ & $\ket{1}_{2}=\ket{b}$ & $\ket{1}_{3}=\ket{V}$
    &&$\ket{1}_{4}=\ket{V}$ & & $\ket{1}_{5}=\ket{V}$ & \\

& & $\ket{0}_{1}=\ket{H}$ & $\ket{0}_{2}=\ket{a}$ & $\ket{0}_{3}=\ket{H}$
    &$\ket{0}_{4}=\ket{a}$&$\ket{0}_{5}=\ket{H}$ & & $\ket{0}_{6}=\ket{H}$&\\
\multirow{-2}{*}{6} & \multirow{-2}{*}{II}
& $\ket{1}_{1}=\ket{V}$ & $\ket{1}_{2}=\ket{b}$ & $\ket{1}_{3}=\ket{V}$
    &$\ket{1}_{4}=\ket{b}$&$\ket{1}_{5}=\ket{V}$ & & $\ket{1}_{6}=\ket{V}$&\\
\bottomrule
\end{tabular}

\caption{
\textbf{Qubit encoding.} The table shows the encoding map between
logical states and photons. Photons are labeled with capital letters A, B, C and
D. Two photons (A and B in the table) are used in setup (I) to encode states up
to $4$ qubits in the polarisation and OAM basis. For setup (II) states up to
$6$-qubits are generated adding two extra photons (C and D) and using an
encoding in path and polarisation.
The states $\ket{H}, \ket{V}, \ket{R}, \ket{L}$ denote the polarisation degree
of freedom while $\ket{+1}$ and $\ket{-1}$ represent the eigenstates of the OAM
with $l=+1$ and $l=-1$, respectively. To identify the two possible paths of the
photons in setup (II) we use the labels $\ket{a}$ and $\ket{b}$.
}
\label{tab:encoding}
\end{table*}

The PAC-model has been adapted to quantum states in \cite{aaronson2007learnability}. Here the learner tries to approximate a function $F_{\rho}: \mathcal{S} \rightarrow [0,1]$ where $F_{\rho}$ is defined as $F_{\rho} (E_i ^{(1)}) = \mathrm{Tr} (E_i ^{(1)} \rho)$. The training set corresponds to a set of $m$ tuples $\{( E_i ^{(1)},F_{\rho} (E_i ^{(1)}))  \}$. Notice that we always take the first element $E_i ^{(1)} $ of each POVM $E_i$. For this reason, in the following, we take $E_i ^{(1)} = E_i$. The POVM measurements $\{E_i\}$ are drawn from an unknown distribution $\mathcal{D}$ and the $F_{\rho} (E_i)$ values are determined experimentally. After processing the training set the learner outputs a hypothesis state $\sigma$. A quantum state is considered to be learned if, with probability $1-\delta$, a training set generated according to the distribution $\mathcal{D}$ can be used to predict with probability $\epsilon$ and accuracy $\gamma$ any other measurement drawn from $\mathcal{D}$:
\begin{equation}\label{eq:learncond}
\Pr_{E\in\mathcal{D}}\left[  \left\vert \operatorname*{Tr}\left(
E\sigma\right)  -\operatorname*{Tr}\left(  E\rho\right)  \right\vert
>\gamma\right]  \leq\varepsilon.
\end{equation} 
A pictorial description of this learning procedure is shown in Fig.~\ref{fig:learnigSchematic}.
Because $\sigma$ is a $2^n \times 2^n$-dimensional matrix we would expect that the number of examples in the training set required to learn $\rho$ also scales exponentially. However, it has been proved \cite{aaronson2007learnability} that the number of examples required to learn $F_{\rho}$ scales linearly with $n$ and inverse polynomially with the relevant error parameters (a full statement of the theorem is given in Methods. In the following we shall refer to theorem as Theorem~\ref{qoccam}). More specifically, fixed the error parameters $\epsilon$, $\gamma$ and $\delta$, we can PAC-learn a quantum state provided:

\begin{equation}\label{eq:mscaling}
m\geq\frac{K}{\gamma^{4}\varepsilon^{2}}\left(  \frac{n}{\gamma^{4}%
\varepsilon^{2}}\log^{2}\frac{1}{\gamma\varepsilon}+\log\frac{1}{\delta
}\right), 
\end{equation}
where $K$ is a constant. This result provides an upper bound on the number of measurements required to learn a quantum state with respect to any probability measure over two-outcome POVM measurements. The value of $K$ is left unbounded but it is critical for applying the theorem in an experimental setting. 

The learning procedure prescribed by Theorem~\ref{qoccam} is simple and it involves finding a hypothesis state $\sigma$ such that $\mathrm{Tr} (E_i \sigma) \approx \mathrm{Tr} (E_i \rho) $ for all $i$. Then, with high probability, that hypothesis will generalise in the sense that $\mathrm{Tr} (E \sigma) \approx \mathrm{Tr} (E \rho) $  for most $E$'s drawn from $\mathcal{D}$. It is then possible to interpret the problem of finding a mixed $n$-qubit state which approximately agrees with the measurements as an optimisation problem. 

The optimisation problem takes as input $m$ POVM measurements described by Hermitian matrices $\{ E_1 , \dots, E_m \}$ and their corresponding measurement outcomes $\{ \mathrm{Tr} (E_1 \rho), \dots , \mathrm{Tr} (E_m \rho) \}$. The goal is to find an Hermitian positive semidefinite matrix $\sigma$ that minimises

\begin{gather}\label{eq:optcond2}
f(\sigma) = \sum_{i=1} ^m ( \operatorname*{Tr}(E_i  \sigma) - \mathrm{Tr} (E_i \rho) )^2    \\
\sigma \succeq 0,  \quad   \operatorname*{Tr}(\sigma) = 1,   \nonumber  
\end{gather}
where by $\sigma \succeq 0$ we denote the positive semidefiniteness of $\sigma$.

The above formulation is a convex program whose solution is known to be computable in polynomial time in the dimension of $\sigma$ using interior point methods \cite{nesterov1994interiori, alizadeh1995interior} or the ellipsoid method \cite{grotschel2012geometric}. However, because the dimension of $\sigma$ scales exponentially with $n$,  the problem of finding the minimum of $f(\sigma)$ is in practice not efficiently computable. This is still compatible with the linear scaling of Theorem~\ref{qoccam} (see Methods) because the results proved in~\cite{aaronson2007learnability} are purely information-theoretic and are concerned only with the sample size $m$. For any given class of quantum states, the question is still open of whether hypothesis states can be produced efficiently. In this context, Rocchetto recently proved that stabiliser states are efficiently PAC-learnable~\cite{rocchetto2017stabiliser}.

Finally, we note that learning a quantum state is not a complete replacement for standard quantum state tomography. The PAC-learning framework of Theorem~\ref{qoccam} tests the predictions over the same distribution of the training set; a good hypothesis state could be arbitrarily far from the true state in the usual trace distance metric, but hard to distinguish from the true state with respect to the given distribution over measurements.

\begin{figure*}[t]
\centering
\includegraphics[scale=0.54]{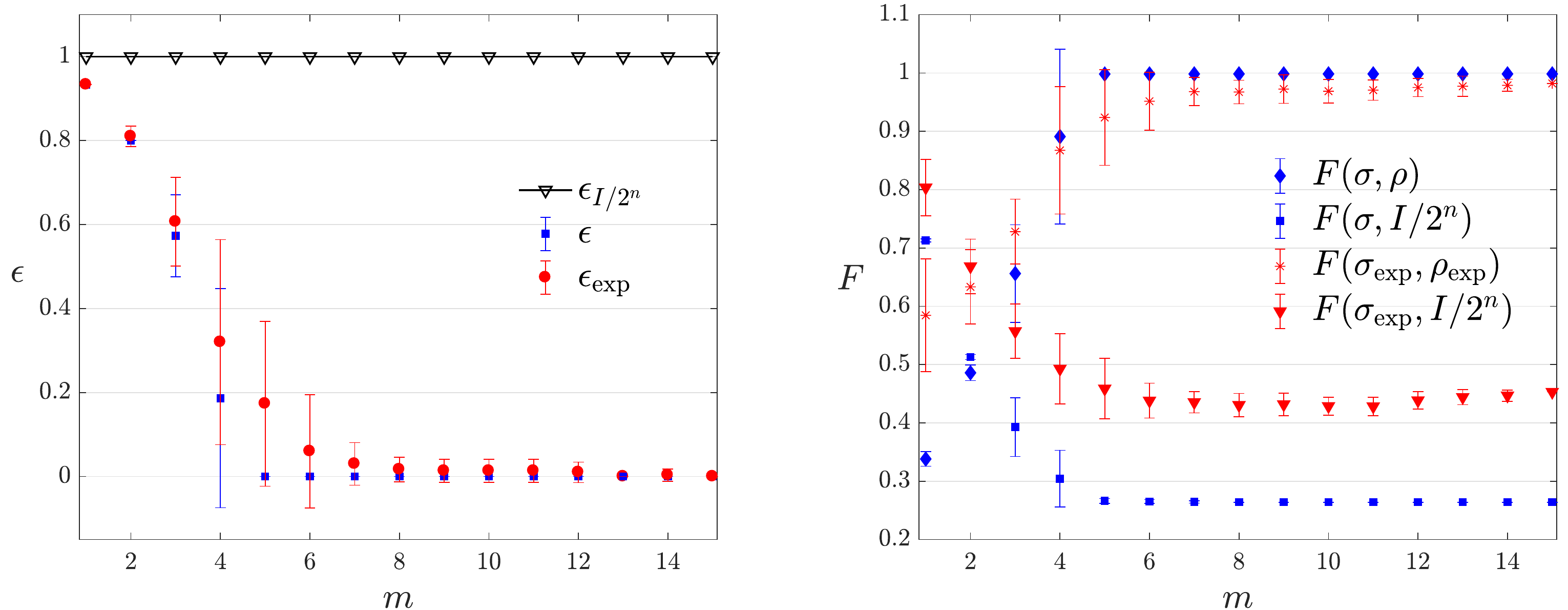}
\caption{\textbf{Learning of a $4$-qubit GHZ state.} Numerical simulations (blue curves) and experimental data (red curves) of the  learning of the state $\left(\ket{0000}+\ket{1111}\right)/\sqrt{2}$. \textbf{(left)} The probability $\epsilon$ of predicting a measurements with less than $\gamma=0.1$ accuracy. The black line represents the predictions made using the completely mixed state as hypothesis. Clearly, the informed predictions are always better than a random guess. \textbf{(right)} The distance, in terms of the fidelity $F = \sqrt{\sigma^{1/2}\rho \sigma^{1/2}}$, between the hypothesis state $\sigma$ and $\rho$ and between $\sigma$ and the completely mixed state $I/2^n$, starting guess of the optimisation algorithm. A discussion on the high variance of the datapoints with $m=4$ is provided in the Methods section. The learning distribution $\mathcal{D}_{(\mathrm{I})}$ is uniform over the set of stabiliser measurements of the state minus the identity matrix (see Methods). Every datapoint is an average of $20$ different, randomly generated, sets of measurement configurations.}
\label{fig:eps_vs_m_dual}
\end{figure*}

\section*{Experimental setup}\label{sec:expset}

We test the learning Theorem~\ref{qoccam} over different Greenberger-Horne-Zeilinger (GHZ) states~\cite{greenberger1989going} (see Methods for a definition). There are several methods to produce GHZ states \cite{walther2006experimental,
huang2011experimental, leibfried2004toward, brattke2001generation, carvacho2015twin, pan_ghz, ciampini_hypentanglement, barreiro, barbi} in photonic
systems. In order to scale up to $6$ qubits we use two different approaches: the first one aims to increase the number of degrees of freedom per photon while the second one exploits an increasing number of photons (see Fig.~\ref{fig:lhv}). 
In setup (I) we generate 2-photon states, encoding up to $4$ qubits, and perform a full set of measurements in the computational basis. In setup (II) we generate four-photon states, able to encode up to $6$ qubits.
Both setups exploit spontaneous parametric down conversion (SPDC) in order to
generate polarisation-entangled photons pairs (see Methods).\\

In setup (I), depicted in Fig.~\ref{fig:lhv}, we use the q-plate \cite{piccirillo2010photon}, a
birefringent patterned slab, to entangle polarisation and orbital
angular momentum (OAM) of single photons \cite{marrucci2006optical, VVbeam, marrucci_stoam_2011, nagali2009quantum}. This makes possible to encode $2$ qubits per photon, exploiting their polarisation and OAM degrees of freedom. 
In order to obtain a $4$-qubit GHZ state, the q-plate acts on the Bell state $\ket{\psi^{-}}=\frac{1}{\sqrt{2}}(\ket{RL}-\ket{LR})$,
where R and L denote, respectively, the right and left circular polarisation of the two photons, allowing
a polarisation--controlled variation of
the OAM. More specifically, states with right or left polarisation become
OAM eigenstates with $\ell = -1$ or $\ell = +1$ respectively.
Conditioned on the measurements of a subset of qubits, we can also generate $3$- and $2$-qubit states, as summarized in table~\ref{tab:encoding}.
In order to perform a complete quantum state tomography in both Hilbert spaces,
the analysis is carried out using two series of quarter-wave plates (QWP), half-wave plates (HWP) and polarising beam
splitters (PBS), separated by another q-plate, to transfer the information
from the OAM to the polarisation
subspace \cite{nagali2009quantum}. The photons are then sent to single mode fibres (SMF), which can be coupled only with states carrying null OAM.

In setup (II), depicted in Fig.~\ref{fig:lhv}, we encode the qubits in the polarisation and path degrees of freedom. Through this encoding we can generate $4$-photon states and up to $6$-qubit.
This setup involves two separate SPDC sources, which generate
two pairs of polarisation-entangled photons, (A,B) and (C,D), with the same pulse of the laser.
We can then obtain a $4$-qubit GHZ state encoded in polarisation, by simultaneously injecting one photon from each source (B and C) over the two inputs of a fibre-based PBS. In this configuration, each photon carries one qubit.
The dimension of the system can be increased to $5$ qubits by sending one of the two output modes of the fibre-based PBS in a Sagnac interferometer (shown in Fig.~\ref{fig:lhv}). This allows us to entangle and measure the polarisation and path degrees of freedom of a single photon while retaining phase stability.
This scheme can be easily extended to $6$ qubits by sending the other output mode of fibre-based PBS in another Sagnac interferometer. In this case, two out of the four photons carry $2$ qubits, which are encoded in the polarisation and path degree of freedom, as shown in Table~\ref{tab:encoding}. 
Through the above procedures we can generate the state
\begin{equation} \label{eq:GHZ6q}
    \ket{\mathrm{GHZ}_n} = \frac{1}{\sqrt{2}}\left(\ket{0}^{\otimes n} + \ket{1}^{\otimes n}\right) 
\end{equation}
for $n = {3, 4, 5, 6}$ qubits.
The polarisation analysis is performed with a HWP and a PBS for each path.\\

\section*{Experimental demonstration}\label{sec:expdem}

\begin{figure*}
\includegraphics[scale=0.6]{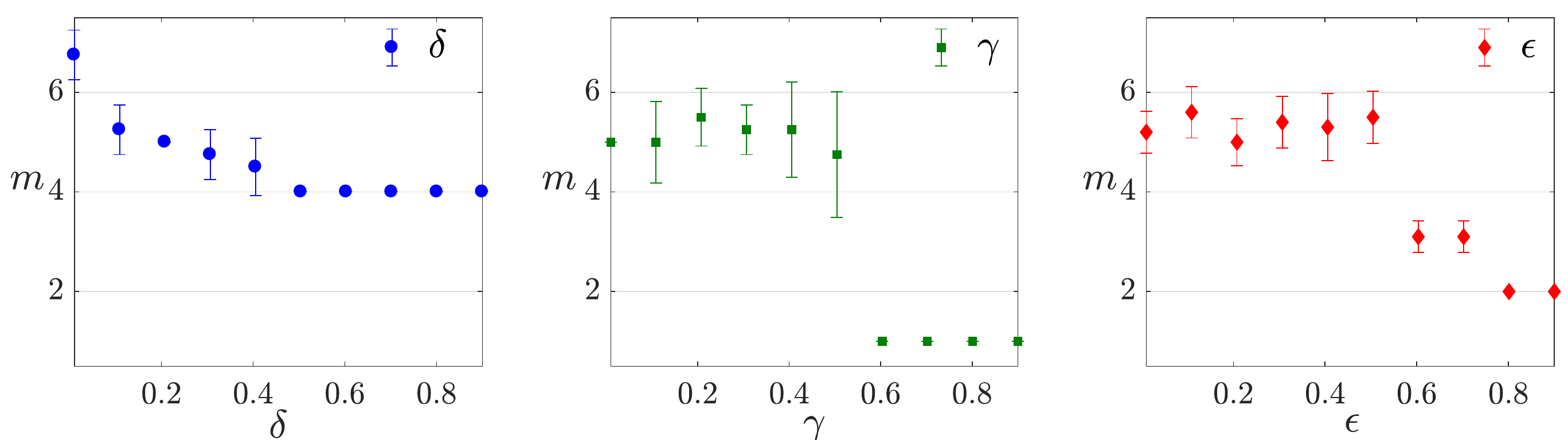}

\caption{\textbf{Measurement complexity of error parameters.} Dependence of $m$
on the error parameters for learning $4$-qubit GHZ states generated with setup (I). Learning is performed under the distribution $\mathcal{D}_{(\mathrm{I})}$ (see Methods for further details) and each data-point is an average over $4$ different GHZ states. When a given error parameter is changed the
other ones are kept constant at the following values $\delta = 0.1$, $\gamma = 0.1$, and
$\epsilon = 0.05$ \textbf{(left)} Scaling of $\delta$. \textbf{(center)} Scaling
of $\gamma$. \textbf{(right)} Scaling of $\epsilon$.}

\label{fig:errorcompl}
\end{figure*}

 We demonstrate numerically and experimentally, through two photonic systems able to encode from $2$ to $6$ qubits, that quantum states can be PAC-learned with a linearly scaling training set:
that is, we demonstrate that the number of elements $m$ in the training set required
to learn an $n$-qubit quantum state $\rho$ scales linearly with $n$.

Although Theorem~\ref{qoccam} can be applied under any distribution
$\mathcal{D}$, it is interesting to test its prediction under distributions that
include measurements that are difficult to predict. If, for example, one were to
take the uniform distribution over all possible measurement bases, with high
probability no measurement drawn from this distribution would be able to
distinguish the state from the completely mixed one. We define the
\textit{completely mixed state} as the state described by the density matrix
$I / 2^{n}$, where by $I$ we denote the identity matrix.

All of our experiments are performed on GHZ states, a type of stabiliser state (see Methods for further details). We remark that the validity of Theorem~\ref{qoccam} extends to all quantum states. The advantage of using GHZ states is the possibility of clearly identifying a set of measurements and a probability distribution that make the predictions of theorem ``interesting'' in the sense that they cannot be reproduced using the completely mixed state as hypothesis. Depending on the experimental setup we use two probability distributions, $\mathcal{D}_{(\mathrm{I})}$ for setup (I) and $\mathcal{D}_{(\mathrm{II})}$ for setup (II), that are uniform over a subset of the
stabilisers of the state (details on the distributions can be found in Methods). Under these
distributions the completely mixed state is never a good hypothesis (unless $\gamma>0.5$) because the
stabiliser measurements performed on the state will always return $1$ as an outcome. On the completely mixed the same measurements will output $1$ or $0$ with equal probability.

In the case of learning with experimental data we have to take into account two
factors that can invalidate Theorem~\ref{qoccam}: noise in the measurements and
the lack of access to the true value of $\mathrm{Tr}(E\rho)$. Both issues can be
positively addressed. We examine the noise problem first. As discussed
in~\cite{aaronson2007learnability}, if the noise that corrupts $E$ to $E'$ is
governed by a known probability distribution such as a Gaussian, then $E'$ is
still just a POVM, so Theorem~\ref{qoccam} applies directly. If the noise is
adversarial, then we can also apply Theorem~\ref{qoccam} directly, provided we
have an upper bound on $ \left\vert \operatorname*{Tr}\left(  E_{i}\rho\right)
-  \operatorname*{Tr}\left(  E'_{i}\rho\right)  \right\vert$. As for the second
issue, approximate values of the expectation values are also within the validity
of the theorem. A discussion is provided in the Methods section.

We begin our experimental analysis with a full characterisation of the PAC-learnability of a $4$-qubit GHZ state generated with setup (I). The complete set of measurements available with setup (I) allows us to compare the quality of the hypothesis $\sigma$ not only in terms of the learning theorem but also from a tomographic perspective. The results, presented in Fig.~\ref{fig:eps_vs_m_dual}, show that, by increasing the number of measurements in the training set, the hypothesis $\sigma$ is getting closer, in terms of fidelity, to the ideal state and to the experimental state (right panel). In the same figure it is possible to see that the predictions (left panel, red dots) obtained by
minimising $f(\sigma)$ are always better than those obtained by taking the
completely mixed state (black line) as hypothesis. This confirms that the distributions we selected are ``interesting'' from a learning perspective because it is not possible to make good predictions using random guessing.

Still using GHZ states generated from setup (I) we test the dependency of the measurement complexity on the error parameters $\epsilon$, $\delta$, $\gamma$. This kind of test is necessary in order to ensure that the hardness of the learning problem used in the experimental demonstration of the theorem is representative of a typical learning scenario. The numerical simulations on the scaling of the error parameters are shown in Fig.~\ref{fig:errorcompl} and indicate that, as expected from Eq.~\ref{eq:mscaling}, the hardness of the learning problem does not change abruptly with the error parameters (unless they introduce pathological cases; for example, for $\gamma>0.5$ random guessing becomes a good prediction strategy).

\begin{figure}
\centering
\includegraphics[scale=0.50]{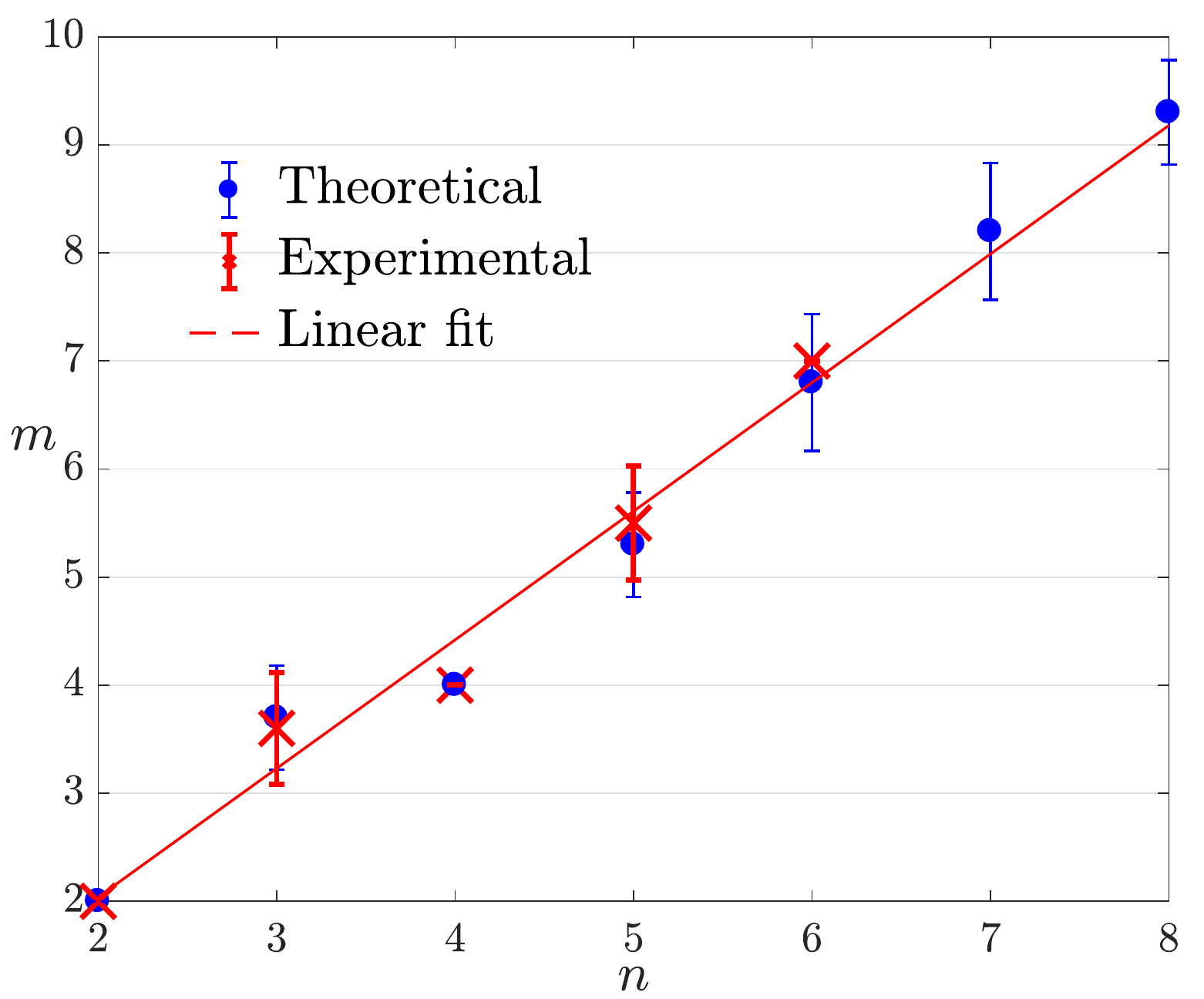}
\caption{\textbf{Experimental demonstration of Theorem~\ref{qoccam}.} Scaling of size of the training set $m$ required to learn a GHZ state as a function of the number of qubits $n$. Experimental data-points (red crosses) are obtained using the experimental setup (II). Each
data-point is obtained using $50$ different, randomly
generated sets of measurement configurations drawn from $\mathcal{D}_{(\mathrm{II})}$ (see Methods for further details). Error bars show the standard deviation for an average of $10$ different runs of the algorithm to estimate $m$. The red line is a linear fit on the experimental data-points with equation $m=1.19n - 0.34$. The learning parameters are
$\epsilon = 0.15$, $\gamma = 0.2$ and $\delta = 0.2$.}
\label{fig:m_vs_n_final}
\end{figure}

We demonstrate the linear scaling of Theorem~\ref{qoccam} over a GHZ of the type described in Eq.~\ref{eq:GHZ6q} and generated by exploiting setup (II). Our
algorithm takes as input the error parameters $ \epsilon,
\gamma, \delta $ and, for a given $n$, outputs the minimum $m$ such that a
training set that respects Eq.~\ref{eq:learncond} is generated with probability
$p = 1 - \delta$.  We present the results in Fig.~\ref{fig:m_vs_n_final} for both numerical and experimental data. The experimental data demonstrates that quantum states are PAC-learnable. A linear fit performed on the experimental data returns a slope value of $1.1$. This implies that the value of the scaling
constant $K$ in Eq.~\ref{eq:mscaling}, left undetermined in
Theorem~\ref{qoccam}, is compatible with learning in an experimental setting. The values obtained from the linear fit in Fig.~\ref{fig:m_vs_n_final} show that learning a $20$-qubit state would require $\sim 23$ measurements. Notice that a
$20$-qubit stabiliser state has $1048576$ stabilisers.

\section*{Discussion}

Our work experimentally demonstrates that quantum states, as a hypothesis class, are PAC-learnable. This result, first proved in \cite{aaronson2007learnability}, constitutes an important advance in our understanding of quantum information. The line of research that seeks to establish how much information is
really contained in a quantum state, and thereby to gain insight about
the reality of the wavefunction, has recently found a new addition in
the ``shadow tomography'' protocol proposed by Aaronson~\cite{aaronson2017shadow}.  This protocol can predict the outcomes of $M$ different two-outcome
measurements on a $D$-dimensional state, to high accuracy, by measuring
only $\mathrm{poly}(\mathrm{log}( D), \mathrm{log}(M))$ copies of the state.
An experimental demonstration of this protocol is a natural future direction, and
would be a valuable addition to our physical comprehension of these
theoretical results.

From a broader perspective, our work constitutes an example of how the techniques developed in the framework of computational learning theory can be used within  quantum information. The interplay of these two fields, recently surveyed by Arunachalam and de Wolf~\cite{arunachalam2017survey}, can offer new tools to investigate properties of quantum states and circuits and can help to identify cases in machine learning where classical and quantum computation behave differently. 
This is particularly important in light of the recent advances in quantum algorithms for machine learning (recently reviewed by Biamonte \textit{et al.}~\cite{biamonte2017quantum} and by Ciliberto \textit{et al.}~\cite{ciliberto2017quantum}) where, despite the growing interest for the topic, it is still unclear whether caveat-free speedups can be attained (for a critical discussion see~\cite{ciliberto2017quantum,aaronson2015read}). 

\subsection*{Acknowledgments}
This work was initiated at the Aspen Center for Physics, which is supported by National Science Foundation grant PHY-$1066293$. The authors would like to thank Simon Benjamin and Ying Li for useful discussions and comments on the manuscript. This work was supported by the ERC-Starting Grant $3$D-QUEST ($3$D-Quantum Integrated Optical Simulation; grant agreement no. $307783$): \href{http://www.3dquest.eu}{http://www.3dquest.eu}. AR is supported by an EPSRC DTP Scholarship and by QinetiQ Ltd. SA is supported by a Vannevar Bush Faculty Fellowship from the US Department of Defense. SS is supported by The Royal Society, EPSRC, the National Natural Science Foundation of China, and the grant ARO-MURI W911NF-17-1-0304 (US DOD, UK MOD and UK EPSRC under the Multidisciplinary University Research Initiative). GC is supported by Becas Chile and Conicyt. The authors would like to acknowledge the use of the University of Oxford Advanced Research Computing (ARC) facility in carrying out this work  \href{http://dx.doi.org/10.5281/zenodo.22558}{http://dx.doi.org/10.5281/zenodo.22558}.


\clearpage

\section*{Methods}

\subsection*{The learning theorem}

The theorem proved in Ref.~\cite{aaronson2007learnability} states:

\begin{theorem}
\label{qoccam}
Let $\rho$ be an $n$-qubit state, let $\mathcal{D}$ be a
distribution over two-outcome measurements, and let $\mathcal{E}=\left(
E_{1},\ldots,E_{m}\right)  $\ consist of $m$ measurements drawn independently
from $\mathcal{D}$. \ Suppose we are given bits $B=\left(  b_1,\ldots
,b_m \right)  $, where each $b_i$\ is $1$ with independent probability
$\operatorname*{Tr}\left(  E_{i}\rho\right)  $ and $0$\ with probability
$1-\operatorname*{Tr}\left(  E_{i}\rho\right)  $. \ Suppose also that we
choose a hypothesis state $\sigma$\ to minimize the quadratic functional
$f(\sigma) = \sum_{i=1}^{m}\left(  \operatorname*{Tr}\left(  E_{i}\sigma\right)
-b_i \right)  ^{2}$. \ Then there exists a positive constant $K$ such that%
\[
\Pr_{E\in\mathcal{D}}\left[  \left\vert \operatorname*{Tr}\left(
E\sigma\right)  -\operatorname*{Tr}\left(  E\rho\right)  \right\vert
>\gamma\right]  \leq\varepsilon
\]
with probability at least $1-\delta$\ over $\mathcal{E}$\ and $B$, provided
that%
\[
m\geq\frac{K}{\gamma^{4}\varepsilon^{2}}\left(  \frac{n}{\gamma^{4}%
\varepsilon^{2}}\log^{2}\frac{1}{\gamma\varepsilon}+\log\frac{1}{\delta
}\right)  .
\]
\end{theorem}

In this article, rather than working with single measurement outcomes $b_i$, we are concerned with estimated expected values  
$$\operatorname*{Tr}\left(E_i \rho \right) \approx \sum_{j=1} ^S b_i ^{(j)}  / S $$
where each $b_{i} ^{(j)}$\ is the $j$-th measurement outcome corresponding to $E_i$. In order to show that the hypothesis $\sigma$ generated by considering the expected values is equivalent to that obtained by taking the measurements outcome $b_i$, we define 
$$f' = \sum_{i=1}^{m'}\left(  \operatorname*{Tr}\left(  E_{i}\sigma\right) - \operatorname*{Tr}\left(E_i \rho \right) \right)  ^{2}.
$$ 
If we take $m = m'S$ and solve for $\sigma$ the equations $df/d\sigma = 0$ and $df'/d\sigma = 0$  it is possible to verify that the hypothesis that minimises the function $f'$ is also satisfying $f$. 

\subsection*{The learning distributions}

We use different learning distributions for the two experimental setups, $\mathcal{D}_{(\mathrm{I})}$ and $\mathcal{D}_{(\mathrm{II})}$. The distribution $\mathcal{D}_{(\mathrm{I})}$ is uniform over the set of stabiliser measurements~\cite{gottesman1996stabilisers} of the GHZ state minus the identity matrix. The distribution $\mathcal{D}_{(\mathrm{II})}$ is uniform over the set of stabiliser measurements in $X$ and $Z$ of the GHZ state minus the identity matrix. A GHZ state \cite{greenberger1989going} is a type of stabiliser state. A stabiliser state $\ket{\psi}$ is the unique eigenstate with eigenvalue $+1$ of a set of $N$ commuting multi-local Pauli operators $P_i$'s. That is, $
P_i \ket{\psi} = \ket{\psi}
$, where $P_i = \bigotimes _j w_j$ and $w_j \in \{I,\sigma^x,\sigma^y,\sigma^z\}$ are the Pauli matrices. We define the $P_i$ as the stabilisers of the state.

There are $2^n$ different stabilisers for an $n$-qubit stabiliser state. Because one of the stabilisers is always the identity (whose eigenvalue is $1$ for every state) we chose not to include this measurement in those sampled by $\mathcal{D}$.

Each $P_i$ is a two-outcome observable (with eigenvalues $+1$ or $-1$). We construct the POVM elements $E_i ^{(1)}$ and $E_i ^{(2)}$ of the observable $P_i$ by noting that $ E_i ^{(1)} + E_i ^{(2)} = I$ and $ E_i ^{(1)} - E_i ^{(2)}= P_i$. The POVM element $E_i ^{(1)}$ can be then written as $  E_i ^{(1)} = (I + P_i)/2$.

The set of stabilisers of a state form a group under the operation of matrix multiplication. To represent a state it is then sufficient to consider the $n$ stabilisers that generate this group. For a $n$-qubit state there are $n$ elements in the set of generators. 

The high variance around $m=4$ in Fig.~\ref{fig:eps_vs_m_dual} can be explained in the following way: each datapoint is obtained by averaging over a number of different configurations sampled from $\mathcal{D}_{(I)}$. It is then likely to sample a configuration that includes $2$ generators and $2$ other stabilisers that can be obtained by the product of the generators. It is easy to see how the information content of such a configuration is less than the one where $4$ independent stabilisers are sampled. This will in turn limit the ability of $\sigma$ to output good predictions and will generate the high variance in the data.

\subsection*{Numerical simulations}

We minimise the function $f$ over the positive semidefinite matrices of unit trace with a variant of the Frank-Wolfe algorithm~\cite{frank1956algorithm} developed by Hazan~\cite{hazan2008sparse}.  All our simulations are performed using $300$ iterations of the Hazan algorithm.

\subsection*{Experimental details}\label{sec:expsetup}

For the experimental setups of Fig.~\ref{fig:lhv}, a pump laser with $\lambda=397.5$ nm is produced by a second harmonic generation (SHG) process from a Ti:Sapphire mode locked laser with repetition rate of $76$ MHz. Photon pairs entangled in the polarisation degree of freedom are generated exploiting type-II SPDC in 2 mm-thick beta-barium borate (BBO) crystals. The photons
generated by SPDC are filtered in wavelength and spatial mode by using narrow
band interference filters and SMF, respectively. After
coupling into SMF, the spatial mode becomes a fundamental Gaussian mode
($TEM_{00}$) with null associated OAM.
 
\clearpage

\appendix

\begin{widetext}
\section{Theorem~\ref{qoccam} with expected measurement values}

Theorem~\ref{qoccam} is stated in terms of the single measurement outcomes $b_i$. Here we show that the results of the theorem still hold if, rather than consider single measurement outcomes, we work with the estimated expected values of $ \operatorname*{Tr}\left(E_i \rho \right) \approx \sum_{j=1} ^s b_i ^{(j)}  / s$ where each $b_{i} ^{(j)}$\ is $1$ with independent probability $\operatorname*{Tr}\left(  E_{i}\rho\right)  $ and $0$\ with probability $1-\operatorname*{Tr}\left(  E_{i}\rho\right)  $. To establish the equivalence it suffices to show that the $\sigma$ that minimises $f  = \sum_{i=1}^m (  \operatorname*{Tr}(  E_{i}\sigma) -b_i)^2$ also minimises $f' = \sum_{i=1}^{m'}\left(  \operatorname*{Tr}\left(  E_{i}\sigma\right) - \operatorname*{Tr}\left(E_i \rho \right) \right)  ^{2}$ where $m=m's$. 

For an integer $s$, let $[s]$ denotes the set $\{1,\dots, s\}$. If we assume that there exist $s$ different measurements $\{b_i ^{(j)}\}_{j\in [s]}$ of each operator $E_i$ we can rewrite $f$ by grouping together measurement outcomes that correspond to a single POVM:
\begin{equation*} 
\begin{split}
\sum_{i=1} ^m \left(  \operatorname*{Tr} (  E_{i}\sigma) - b_i \right)^2  & = ( \operatorname*{Tr} (  E_{1}\sigma) - b_1 ^{(1)} )^2 +  ( \operatorname*{Tr}(E_{1}\sigma) - b_1 ^{(2)} )^2  + \dots + (\operatorname*{Tr}(  E_{m'}\sigma) - b_{m'} ^{(s-1)} )^2 + (\operatorname*{Tr}(  E_{m'}\sigma) - b_{m'} ^{(s)} )^2 \\
&= \sum_{i=1}^{m'} \left[ s \left(  \operatorname*{Tr}(  E_{i}\sigma) \right)^2 +  \sum_{j=1} ^s (b_i ^{(j)})^2 - 2\operatorname*{Tr}(  E_{i}\sigma ) \sum _{j=1} ^s b_i ^{(j)} \right] \\
 & =  \sum_{i=1}^{m'}  s \left[ \left(  \operatorname*{Tr}(  E_{i}\sigma) \right)^2 +  \sum_{j=1} ^s (b_i ^{(j)})^2 /s - 2\operatorname*{Tr}(  E_{i}\sigma) \sum_{j=1} ^s b_i ^{(j)} /s \right].
\end{split} 
\end{equation*}
Equivalently $f'$ can be expressed as:
\begin{equation*} 
\begin{split}
\sum_{i=1}^{m'}\left(  \operatorname*{Tr}\left(  E_{i}\sigma\right) -\operatorname*{Tr}\left(  E_{i}\rho\right) \right)  ^{2} & = \sum_{i=1}^{m'}\left(  \operatorname*{Tr}\left(  E_{i}\sigma\right) -\sum_{j=1} ^s b_i ^{(j)}  / s \right)  ^{2}\\
 & = \sum_{i=1}^{m'}  \left[ \left(  \operatorname*{Tr}\left(  E_{i}\sigma\right) \right)^2 +  \left( \sum_{j=1} ^s b_i ^{(j)}/s \right)^2- 2 \operatorname*{Tr}\left(  E_{i}\sigma\right) \sum_{j=1} ^s b_i ^{(j)} /s \right].
\end{split}
\end{equation*}
The minimum of $f(\sigma)$ is found for: 
\begin{equation*} 
\frac{df(\sigma)}{d\sigma} = \sum_{i=1}^{m'}  \left[ \frac{ d\operatorname*{Tr}(  E_{i}\sigma) )^2}{d\sigma} - 2\frac{d\operatorname*{Tr}(  E_{i}\sigma)}{d\sigma} \sum_{j=1} ^s b_i^{(j)} /s \right] = 0
\end{equation*}
Equivalently, we get for $f'$:
\begin{equation*} 
\frac{df'(\sigma)}{d\sigma} = \sum_{i=1}^{m'}  \left[  \frac{d\operatorname*{Tr}\left(  E_{i}\sigma\right)^2}{d\sigma}- 2 \frac{d\operatorname*{Tr}\left(  E_{i}\sigma\right)}{d\sigma} \sum_{j=1} ^s b_i^{(j)} /s \right] = 0.
\end{equation*}
It is easy to see how $f$ and $f'$ are minimised by the same $\sigma$.

\section{Algorithm to estimate the scaling of $m$}

With algorithm~\ref{algo:estimatem} we estimate the minimum number of measurements $m$ that allows us to PAC-learn $\rho$ with accuracy parameters $\epsilon$, $\gamma$ and success probability $1-\delta$. For each iteration of $i$ the algorithm generates a set of measurements drawn from either $\mathcal{D}_{(I)}$ or $\mathcal{D}_{(II)}$. We give the pseudocode for the case of $\mathcal{D}_{(I)}$. The support of $\mathcal{D}_{(I)}$ is the set $\mathcal{V}$ of stabiliser measurements of the state minus the identity operator. Because each stabiliser state has $2^n$ stabiliser measurements we have  $|\mathcal{V}| = 2^{n}-1$. 

The case for $\mathcal{D}_{(II)}$ is identical apart for  the support of $\mathcal{D}_{(II)}$ that is now  the set $\mathcal{W}$ of the stabiliser measurements on $X$ and $Z$ of the state minus the identity operator. 

\alglanguage{pseudocode}
\begin{algorithm}[H]
\caption{Find minimum $m$ that allows to PAC-learn $\rho$}
\label{algo:estimatem}
\textbf{Input:} quantum state $\rho$, number of qubits $n$, distribution $\mathcal{D}_{(I)}$, error parameters $\epsilon, \gamma, \delta $, number of different training sets used for the estimate $i_{\mathrm{MAX}}$ \\
\textbf{Output:} minimum value of $m$ that satisfies the conditions of Theorem~\ref{qoccam} \\
\begin{algorithmic}[1]
\State $m = 1$
\Repeat
\State $\delta_{est} =  0 $
\For{$i=1 \dots i_{MAX}$}
\State Generate training set $T=\{(E_i,\mathrm{Tr}(E_i\rho))\}_{i\in[m]}$ with random measurements drawn from $\mathcal{D}_{(I)}$
\State $\sigma = \mathrm{HAZAN}(T,n)$
		\For{every $E\in \mathcal{V}$}
			\If{$| \mathrm{Tr}(E \sigma) - \mathrm{Tr}(E_i \rho)| > \gamma$}
				\State $\epsilon_{\mathrm{est}}  += 1/ |\mathcal{V}|$
			\EndIf
		\EndFor
	\If{ $\epsilon_{\mathrm{est}}> \epsilon$}
	\State $\delta_{est} += 1/i_{\mathrm{MAX}}$
	\EndIf
\EndFor
\State $m = m + 1$
\Until{$\delta_{est} < \delta$}
\end{algorithmic}
\end{algorithm}

\section{The Hazan's algorithm}

As discussed the problem of learning quantum states can be cast as a convex program. In the formulation given in Eq.~\ref{eq:optcond2} the goal is to minimise the objective function $f(\sigma) =  \sum^m _{i=1} (\mathrm{Tr}(E_i\sigma) - \mathrm{Tr}(E_i\rho))^2$ over the positive semidefinite matrices of unit trace. Because both the space of positive semidefinite matrices of unit trace and the objective function are convex, we are dealing with a constrained convex optimisation problem. A polynomial time algorithm for this class of problems is the Frank-Wolfe algorithm \cite{frank1956algorithm} for optimising a single function over the bounded positive semidefinite cone. In our simulations we use an extension of this work, developed by Elad Hazan \cite{hazan2008sparse}, specifically designed for learning quantum states with the procedure described in Theorem~\ref{qoccam}.

\alglanguage{pascal}
\begin{algorithm}[H]
\label{Hazan}
\caption{Hazan's algorithm}
\textbf{Input:} training set $T=\{(E_i,\mathrm{Tr}(E_i\rho))\}_{i\in[m]}$, Hilbert space dimension $N=2^n$, and maximum number of iterations $k_{\mathrm{MAX}}$  \\
\textbf{Output:} hypothesis state $\sigma$ \\
\begin{algorithmic}[1]
\State Initialise $\sigma_0 = I/N$
\For{k=1 }{k_{\mathrm{MAX}}}
\Begin
\State Compute the smallest eigenvector $v_k$ of $\nabla f(\sigma_k)$
\State Let $\alpha = \frac{1}{k}$
\State Update $\sigma_{k+1} = \sigma_k + \alpha_k(v_k v_k ^T - \sigma_k)$
\End
\end{algorithmic}
\end{algorithm}

We can compute analytically step $4$ by using that $\frac{\partial \mathrm{Tr}(F(\mathbf{X}))}{\partial \mathbf{X}} = f(\mathbf{X})^T$, where $f$ is the scalar derivative of $F$, and the hermiticity of the measurement operators $E_i$:

\begin{equation*} 
\begin{split}
\nabla f(\sigma_k) & = \frac{\partial f(\sigma_k) }{\partial \sigma_k} \\
 & = 2 \sum_{i=1} ^m (\mathrm{Tr}(E_i \sigma_k) - \mathrm{Tr}(E_i\rho))E_i^T \\
 & = 2 \sum_{i=1} ^m (\mathrm{Tr}(E_i \sigma_k) - \mathrm{Tr}(E_i\rho))E_i
\end{split}
\end{equation*}

\end{widetext}

\end{document}